\documentclass[letterpaper,twoside,notoc]{JHEP3}

\usepackage{amsmath}
\usepackage{mathrsfs}
\usepackage{graphicx}

\newcommand{\gev}{{\ \rm GeV}}
\newcommand{\tev}{{\ \rm TeV}}

\newskip\zatskip \zatskip=0pt plus0pt minus0pt
\def\matth{\mathsurround=0pt}
\def\lsim{\mathrel{\mathpalette\atversim<}}

\def\atversim#1#2{\lower0.7ex\vbox{\baselineskip\zatskip\lineskip\zatskip
  \lineskiplimit 0pt\ialign{$\matth#1\hfil##\hfil$\crcr#2\crcr\sim\crcr}}}

\author{JoAnne L. Hewett$^a$, Ben Lillie$^a$, 
Manuel Masip$^{a,b}$ and Thomas G. Rizzo$^a$\thanks{Work supported by the 
  Department of Energy
  Contract DE-AC02-76SF00515. M.M. also supported by a grand from the
  MECYD (PR2003-0466), by MCYT (FPA2003-09298) and Junta de Andaluc{\'\i}a
  (FQM-101).} \\ $^a$Stanford Linear Accelerator Center, 2575 Sand Hill
  Rd. Menlo Park, CA 94025\\ $^b$CAFPE and Departamento de F{\'\i}sica 
 Te\'orica y del Cosmos\\
Universidad de Granada, 18071, Granada, Spain\\
E-mail: \email{hewett@slac.stanford.edu,
  lillieb@slac.stanford.edu, masip@slac.stanford.edu, rizzo@slac.stanford.edu}}

\title{Signatures of long-lived gluinos in split supersymmetry}
\preprint{SLAC-PUB-10630\\UG-FT-166-04}

\abstract{
We examine the experimental signatures for the production of gluinos at 
colliders and in cosmic rays within the split supersymmetry scenario. 
Unlike in the MSSM, the gluinos in this model are relatively long-lived 
due to the large value of the squark masses which mediate their decay. 
Searches at colliders are found to be sensitive to the nature of 
gluino fragmentation as well as the gluino-hadron interactions with nuclei 
and energy deposition as it traverses the detector.  We find that the 
worst-case scenario, where a neutral gluino-hadron passes through the 
detector with little energy deposition, is well described by a monojet
signature.   For this case, using Run I data we obtain a bound of 
$m_{\tilde g} > 170$ GeV;  this will increase to 210(1100) GeV at 
Run II(LHC) if no excess events are observed.  In the opposite case, where 
a charged gluino-hadron travels through the detector, a significantly 
greater reach is obtained via stable charged particle search techniques.  
We also examine the production of gluino pairs in the atmosphere by 
cosmic rays and show they are potentially observable at IceCube; this
would provide a cross-check for observations at hadron colliders.
}

\begin{document}

\section{Introduction}\label{sec:introduction}

The Minimal Supersymmetric Standard Model (MSSM) enjoys considerable
phenomenological success, including the generation of electroweak
symmetry breaking,  the prediction of grand
unification, and the presence of a dark matter candidate
\cite{Ellis:2002mx}.
However, the existence of naturally large supersymmetric contributions 
to flavor changing neutral currents \cite{Ellis:1981ts}, 
to electric dipole moments (edm) \cite{Ellis:1982tk,Polchinski:1983zd}, 
and to proton decay are well-known and have long plagued supersymmetric
model builders.
In addition, the lack of observation of a light Higgs boson at LEP II
\cite{LEPII:2001xx} has created tension in the MSSM parameter space and 
indicates the existence of tuning at the level of at least a few percent.
These outstanding issues have prompted a continuing vast array of research
over the last two decades, and numerous candidate supersymmetric models
have been put forward.

Recently, a supersymmetric model has been proposed 
\cite{Arkani-Hamed:2004fb,Giudice:2004tc}  which retains the
successes of supersymmetry, solves the flavor and CP problems, and extends
the proton lifetime, albeit at the cost of naturalness.  This model,
known as Split Supersymmetry, postulates that Supersymmetry breaking
occurs at the very high scale, $m_S \gg 1000 \tev$.  The scalar
particles all acquire masses at this high scale, except for a single 
neutral Higgs boson, whose mass is either finely-tuned or is preserved by
some other mechanism.  The fermions in the theory, including the gauginos, 
are assumed to be protected by chiral symmetries and thus can have 
weak-scale masses.  (For other supersymmetric models with heavy
scalars and weak-scale fermions, see \cite{Kumar:2004yi}.)
The existence of weak-scale fermions ($i$) preserves 
gauge coupling unification even if the scalar partners are ultra-heavy 
since only complete $SU(5)$
multiplets are being removed from the low energy spectrum, and ($ii$) 
provides a natural dark matter candidate in the lightest neutralino.
The ultra-heavy masses for the scalars guarantee the absence of large 
supersymmetric flavor changing interactions, since all such processes 
are mediated by the sfermions at loop level if R-parity is conserved
(or at tree-level if R-parity is violated).  The generic constraints from
flavor and edm data which set $m_S > 100 \tev$ and $1000 \tev$, respectively,
are easily satisfied within this model.  Likewise, the familiar 
dimension-five operator which mediates proton decay, $qq\tilde q\tilde\ell$,
is also suppressed, delaying proton decay which now occurs via 
dimension-six operators.  This increase in the proton lifetime 
is also in agreement with data\cite{Murayama:2001ur}.  Since the theory
is not supersymmetric below the scale $m_S$, the usual relations between the
supersymmetric gauge and Higgs sector ({\it i.e.}, between
the Yukawa, quartic and gauge couplings) no longer hold.  In particular, 
Renormalization Group
Evolution in this model yields a prediction \cite{Arvanitaki:2004eu} for 
the lightest Higgs mass of $m_h = 130-170 \gev$; this 
differs from that of the MSSM and is in agreement with the null
searches at LEPII.  However, since the value of the sfermion masses
are ultra-heavy, the large quadratic corrections to the mass of the Higgs
are not cancelled in the manner present in weak-scale supersymmetry, and
the Higgs sector remains extremely fine-tuned.  Split Supersymmetry
proponents argue that this tuning may, indeed, be present in nature
and may be related to the cosmological constant problem which suffers
an even greater degree of fine-tuning.  For example, it is possible that 
the tuning in the Higgs mass may be consistent with the
solution to the cosmological
constant problem based on the string theory landscape. 

Whether or not
one takes the motivation for Split Supersymmetry seriously, it is important 
to consider the phenomenological implications of having such a 
fine-tuning present in nature.
One generic feature of this model is that the gluino is very long-lived, 
since the squarks which mediate its decay are ultra-heavy.  It could 
easily appear to be stable in collider processes.
In this paper, we will examine the signatures of a long-lived
gluino in both collider and cosmic ray detectors.  Collider
signals for such particles have been previously examined in
\cite{Baer:1998pg,Raby:1998xr,Mafi:1999dg},
and are found to be dependent on the charge of the gluino-hadron
resulting from fragmentation which dictates the amount of energy deposition 
by the gluino hadron as it traverses the detector.  Here, 
we point out the importance of the monojet signature, arising from
gluino pair production plus a jet bremsstrahlung with the gluinos escaping
the detector unobserved;  this signal is present for
all possible charges of the gluino-hadron.  We also consider the
charged gluino-hadron constraints from charged stable particles searches.
In both cases, we find constraints from current data from $p\bar p$ 
collisions at the Tevatron Run I, and then estimate the search reach of 
Run II and the LHC.  In addition, we examine the signals in very large 
neutrino detectors, such as IceCube, from the production of gluino pairs
in the atmosphere by cosmic rays.  All our results are essentially 
independent of the Split Supersymmetry model details and are applicable 
to any model which contains a stable or meta-stable heavy colored particle.

This paper is organized as follows. In section \ref{sec:splitsusy} we
briefly review Split Supersymmetry, and the properties of the
gluino in this proposal. In section \ref{sec:searches} we show how
constraints can be placed on the mass of the gluino from present Tevatron
data, and what the search reaches might be at Run II and the LHC. In
section \ref{sec:cosmic} we describe a possible signal in the IceCube
detector from the cosmic ray production of gluinos.
Section \ref{sec:conclusion} concludes.

\section{Phenomenological Features of Split Supersymmetry and Long-Lived
Gluinos}\label{sec:splitsusy}

The rationale behind Split Supersymmetry has been discussed in the previous
section.  The essential phenomenological ingredients are that supersymmetry
breaking occurs near the GUT scale with $m_S\gg 10^6 \tev$ and all
scalar masses, except for a single finely-tuned light Higgs, are set 
to that scale.  Whereas the fermions are
protected by a chiral symmetry and have masses at the weak scale.
A feature of this supersymmetric mass spectrum is the dramatic reduction 
over the MSSM in the number of parameters at the TeV scale.
In particular, there are only 9 new
parameters relevant at TeV energies. These are the three gaugino masses, 
a $\mu$ term for the Higgsinos, the four Higgsino-Gaugino couplings, and 
the scale of the scalar masses and supersymmetry breaking, $m_S$. Many 
observables will only depend on the
gaugino masses and $m_S$, yielding predictions which are
robust with respect to variation over the parameter space.  We also note 
that since supersymmetry is not present at the TeV scale, the weak-scale
Higgsino-Gaugino couplings differ from their usual MSSM values;
these couplings approach these values as one evolves towards $m_S$.

This scenario produces a strikingly different phenomenology from that of 
the MSSM.  Some features have been examined in recent papers, including the 
implications for dark matter detection \cite{Giudice:2004tc,Pierce:2004mk},
the Renormalization Group running from the high scale $m_S$ 
\cite{Arvanitaki:2004eu}, and aspects of sparticle detection at colliders
\cite{Giudice:2004tc,Zhu:2004ei,Mukhopadhyaya:2004cs}.

The absence of the scalars at the TeV scale affects both the production
and decays of the gauginos.  Although we will focus on the gluinos in 
this paper, we first make a few comments on the electroweak gaugino
sector.  In the MSSM, most
sparticles decay through a cascade down to the lightest supersymmetric
particle (LSP), and these cascade decays usually involve sfermions at 
some stage. Due to the heavy scalar masses in Split Supersymmetry, these 
cascade decays will essentially not occur.  This has a
two-fold effect on the electroweak gaugino phenomenology:  ($i$)  
Cascade decays of heavier sparticles provide the main production 
mechanism for charginos and neutralinos at hadron colliders.  Since
the squarks are beyond the kinematic reach and, as we discuss at length below, 
the gluino is long-lived, cascade decays of these particles are no
longer a source of electroweak gaugino production.  In addition, heavier
gauginos are not likely to cascade.  Hence the electroweak gauginos
can only be produced via the Drell-Yan mechanism.  ($ii$)  The
only decay channels open to the electroweak gauginos are direct
decays to the electroweak gauge bosons, 
such as $\tilde\chi^\pm\to W^\pm+\tilde\chi^0_1$.  The characteristic
tri-lepton signature for electroweak gaugino production is still
viable, but the lepton spectrums will be modified.
Combined, these two features result in a lower production rate
for charginos and neutralinos at hadron colliders and a more 
prompt lepton spectrum arising from their decays. 

The lack of TeV scalars also affects the production and decay of
the gluinos.  The usual Feynman diagrams for gluino pair production
via $gg$ and $q\bar q$ annihilation in the MSSM are displayed in 
Fig. \ref{fig:gpairfeyn}.  The fourth diagram proceeds by squark
exchange and while it can make a significant contribution to the
production cross section in the MSSM, it 
is negligible in the Split Supersymmetry model.
The channels for gluino pair production in this scenario are thus
the same as for heavy quark production in the Standard Model (SM).
In addition, the associated production channel $q\bar q\to\tilde g+
\chi^0_i/\chi^\pm_i$ is important in the MSSM as it can either
contribute to the jets$+\slash \!\!\!\! E_T$ signal or produce leptons
in the final state.  This production proceeds via $\tilde q_{L,R}$
exchange in the $t$- and $u-$ channels, and does not occur
in the Split Supersymmetry scenario.  In summary, we see that the
absence of TeV scalars results in production
cross-sections for gluinos that are somewhat lower than those 
generally considered in the MSSM.

We note that searches at LEP for stable hadronizing
gluinos have been carried out \cite{Abdallah:2002qi,Heister:2003hc}.  
Both ALEPH and DELPHI
have searched in the channel $e^+e^-\to q\bar qg\to q\bar q\tilde
g\tilde g$, and place the constraint $m_{\tilde g}>26.9$ GeV
at 95\% CL.  They have obtained stronger limits in the case of
squark pair production with subsequent decay into long-lived
gluinos, however, these channels do not occur in the present 
scenario.

\FIGURE[t]{
\includegraphics[,width=12cm]{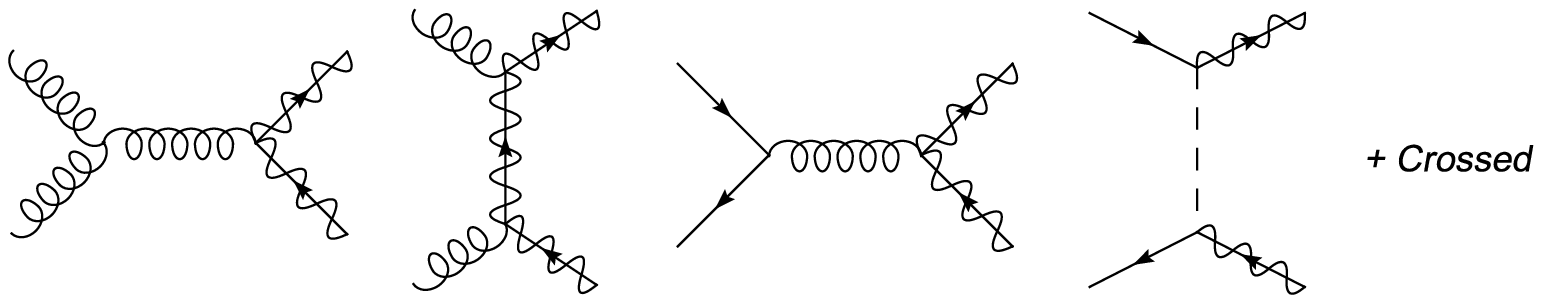}
\label{fig:gpairfeyn}
\caption{Diagrams contributing to gluino pair production in hadronic
  collisions in a generic supersymmetric model. The fourth diagram
  proceeds by squark exchange, and is effectively absent in Split
  Supersymmetry. The remaining diagrams are identical to those which
  mediate heavy flavor pair production.}}

A striking feature of the Split Supersymmetry model
is the extremely long lifetime of the gluino. 
Unlike the electroweak gauginos, it does not have a decay channel
that avoids sfermions. The gluino can only decay through a virtual squark to
a $q\bar q(q') + \chi^0_i(\chi^\pm_i)$. 
The large mass of the squarks then implies a
long lifetime for the gluino. This lifetime is given approximately by
(neglecting the mass of the electroweak gaugino)\cite{Dawson:1983fw}
\begin{gather}
\tau \simeq 8\left( \frac{m_S}{10^9 \gev}\right)^4
\left(\frac{1 \tev}{m_{\tilde g}}\right)^5 {\rm s}\,.
\end{gather}
Here we have explicitly included the coupling constants and summed over the
possible final states.
Since the scale $m_S$ ranges from roughly $10^7 \gev$ up to the 
GUT scale (although $m_S< 10^{12-13}$  is favored for
cosmological reasons \cite{Arkani-Hamed:2004fb}), the gluino lifetime
can easily vary from picoseconds to the age of the universe for
gluino masses near the electroweak scale.  If the
lifetime is too long, then there may be cosmological problems from the 
relic density of gluinos. There is a strong limit on the density 
of heavy isotopes \cite{Smith:1982qu},
which implies that either the gluino lifetime 
must be smaller than the age of the
universe, or there is some mechanism in the early universe that insures
they have a vanishingly small relic density. 

A long-lived, but unstable, gluino produces interesting signals in
collider experiments.
For lifetimes around $1 {\rm ps}$ the gluino will most likely decay
inside a silicon vertex detector, and could be found in a search for
multi-jets plus $\slash \!\!\!\! E_T$ with displaced vertices. For lifetimes
between $\sim 1 {\rm ps}$ and $\sim 100 {\rm ns}$, the gluino will decay in
the bulk of the detector.  However, 
the signals from this decay will most likely be lost in the
background, or not pass the trigger.  It may be possible to
reconstruct them off-line if the event contained something else which
passed the trigger.  Generically, however, 
for the bulk of the parameter space, the
lifetime is expected to be larger than $10^{-7} {\rm s}$, and
gluinos produced in colliders will decay outside of the detector. In this
case they will appear to be effectively stable, and search strategies
for heavy stable particles need to be employed \cite{Perl:2001xi}. 

When the gluino is produced, it will hadronize into a
color singlet state, called an $R$-hadron (since it carries one unit of
$R$-parity)\cite{Farrar:1978xj,Farrar:1978rk}.  For notational purposes,
we will denote the hadron $\tilde gq\bar q$ as an $R$-meson, $\tilde gqqq$ as
a $R$-baryon and $\tilde gg$ as a $R$-gluon.
Search techniques for such particles will depend on the
characteristics of the $R$-hadron, in particular its electric charge.
We discuss each possibility in detail below.

\section{Searches at Hadron Colliders}\label{sec:searches}

\subsection{Propagation inside a medium}\label{subsec:prop}

Once produced, the long-lived gluino will hadronize into a
color singlet state $R$. If neutral, $R$ will experience
energy loss only through hadronic collisions as it propagates 
through matter, whereas a charged $R$ will also deposit
energy in the form of ionization.  Notice that there is also
the possibility that $R$ can change its charge in each hadronic interaction,
resulting in a ``flipper'' R-hadron that is alternately charged
and neutral as it propagates through matter.  

The fragmentation
probabilities into neutral versus charged R states are extremely
uncertain.  However, there are two cases which favor fragmentation 
to a neutral $R$-hadron:  {\it (i)} the 
probability of fragmentation to $\tilde g qqq$ is much smaller 
than to $\tilde g g$, and {\it (ii)} 
the mass difference between $\tilde g q \overline q$ and the 
neutral state $\tilde g g$ is larger than $m_\pi$ ($R$-meson
states would then decay quickly into $\tilde g g$ and the produced 
hadron would be effectively always neutral).
In any other case one would naively expect similar probabilities for
the hadronization into charged or neutral states. 

The possibility of a charge exchange for a gluino hadron following hadronic
scattering is more involved. In
particular, the fact that $R$-meson (and $R$-gluon) states may 
convert into baryon $R$ hadrons 
({\it e.g.}, $\tilde g d\overline d + uud \rightarrow 
\tilde g udd + u\overline d)$ but not vice-versa \cite{Kraan:2004tz} 
has been overlooked in previous analyses of long-lived gluinos 
\cite{Baer:1998pg,Mafi:1999dg}. This conversion
would be favored by the lightness of the final pion, whereas
processes that convert $R$-baryon into $R$-meson states
are negligible. In any case, to cover all the possibilities
we will consider the limits where $R$ is always
neutral or always charged, and will also briefly discuss the 
flipper model proposed in \cite{Kraan:2004tz}.

As $R$ propagates, hadronic energy loss will be 
dominated by interactions with the nucleons in the medium 
(rather than partons or the nucleus as a whole).  Little is known
about these interactions, however several models have been discussed
in the literature \cite{Baer:1998pg,Raby:1998xr,Mafi:1999dg,Kraan:2004tz}.
The rate of energy loss will depend on the average amount of energy 
deposited in each interaction ($\langle \Delta E \rangle$) 
and the average distance between interactions in that 
medium ($\lambda_T (R)$). The hadron
$R$ can be viewed as a non-interacting heavy gluino surrounded 
by a colored cloud of light constituents that are responsible 
for the interaction. This simple picture suggests
that $\langle \Delta E \rangle$ is 
independent of the gluino mass and only depends on
the speed ($\beta$) of the hadron.  In addition, only a small
fraction of the energy of $R$ is relevant to the interaction.
For example, it is expected that 
a $\tilde g q\overline q$ hadron with $E=400$ GeV and 
$m=200$ GeV would strike a nucleon 
as if it were a light meson of mass $0.6 \gev$ with $E=1.2$ GeV. Therefore,
$R$ will lose a very small fraction of its 
energy in each interaction. 

\FIGURE[t]{\includegraphics[width=10cm,angle=-90]{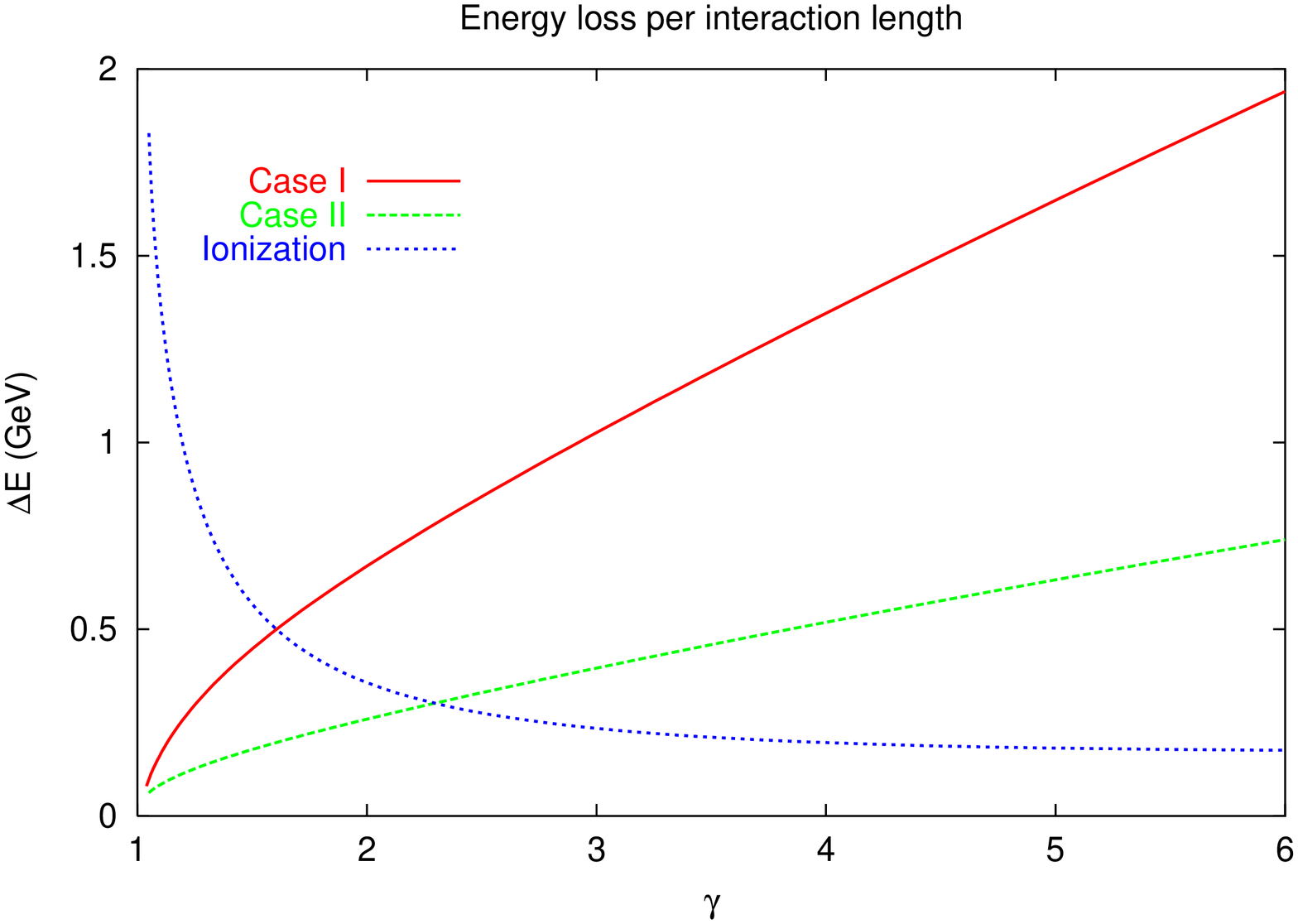}
  \label{fig:energyloss}\caption{Energy loss by a $R$-hadron per
    interaction length due to hadronic interactions described by
    two models and due to electromagnetic ionization.}}

In the process $R N\rightarrow R X$, where $X$ is a system
of one or several final particles and $N$ denotes the incident
nucleon, the energy loss is given by
\begin{gather}
\Delta E= { \frac{m_X^2-m_N^2+ | t |}{2m_N} }\;,
\end{gather}
where the dominant values of the momentum transfer $| t |$ are
of order (0.1-1 GeV)$^2$. 
We follow \cite{Baer:1998pg} and consider two different models
describing the differential cross section $d\sigma/d|t|dm_X$ for
this reaction:  Case {\it (1)} the cross section
is given by a constant differential 
for $| t | < 1\;{\rm GeV}^2$, and Case {\it (2)} it is given by
a triple-Pomeron distribution described in \cite{Moshe:1977fe}.
It is remarkable that in both cases 
$\langle \Delta E \rangle$ can be well approximated 
by a linear expression  when expressed in terms of
$\gamma=(1-\beta^2)^{-1/2}$ (see Fig. \ref{fig:energyloss}).
In particular, for $\gamma\ge 1.3$ 
we find $\langle \Delta E \rangle\approx k \gamma$, 
with $k=0.35$ GeV (case {\it (1)}) or $k=0.14$ GeV 
(case {\it (2)}). These
results are consistent with those of 
\cite{Kraan:2004tz}, which also exhibit an approximate linear 
behavior, where when nuclear effects are included, it is found that 
$k$ lies between $0.3$ GeV for
interactions in Iron and $0.2$ GeV in
Hydrogen. Since we are considering interactions inside a detector
(as well as in the atmosphere in the following section), we
therefore take cases {\it (1)} and {\it (2)} as 
upper and lower estimates for $\langle \Delta E \rangle$.

To estimate the mean interaction length of the $R$-hadron
inside a collider detector, we follow \cite{Baer:1998pg} and take
$\lambda_T (R)=(16/9) \lambda_T (\pi) \approx 19$ cm in iron. The instrumented thickness of the 
calorimeter is 
around $8 \lambda_T (\pi)$ at CDF or $11 \lambda_T (\pi)$
at D0. For a neutral $R$ and 
$\lambda_T (R)=19$ cm this implies an average
of 4.5 hadronic interactions depositing a total energy between 
$1.6 \gamma$ GeV (case {\it (1)}) and $0.63 \gamma$ GeV in CDF. 
For a $R$-hadron mass of $100$ GeV with $E=400$ GeV, this gives a total
energy deposition up to 6.4 GeV.   If $\lambda_T (R)$ 
is reduced by a factor of $1/2$
(which doubles the number of hadronic interactions in
the calorimeter) the maximum energy loss visible at 
CDF would be just 12.4 GeV. Notice that if $m>100$ GeV or
$E<400$ GeV (the typical values to be considered
at the Tevatron) the total energy deposition would be even smaller.

For a charged $R$-hadron, the energy loss through ionization is described
by the Bethe-Bloch equation and becomes 
relevant at low values of $\beta$.
Taking into account the $1/\beta^2$ dependence, we find that energy
deposition by
ionization dominates for $\gamma\le 1.5$ (see Fig. \ref{fig:energyloss}) and
is much smaller than that for hadronic interactions when $\gamma\ge 2$.
For a fast moving charged $R$-hadron (see Fig.~\ref{fig:speed} for the
average speed of a gluino produced at the Tevatron),
around $1.3$ GeV of ionization energy would be deposited in the CDF
detector.  Charged $R$-hadrons of mass
$m\approx 100$ GeV produced with $\gamma\le 1.1$
will lose all their kinetic energy (around 10 GeV) through ionization  
in the hadronic calorimeter.

In summary, fast $R$-hadrons
lose energy mainly through hadronic interactions (their electric charge 
becomes irrelevant), whereas slow charged $R$-hadrons have sizeable 
energy loss only through ionization. In either
case, the total amount of energy deposited in 1 meter of iron (the
instrumented length in a hadronic calorimeter) will be smaller than
$\approx$ 15 GeV and may escape the experimental triggers.  We note
that the specific detector characteristics quoted here are for the
Run I configurations; we assume they are roughly equivalent for Run II.

\FIGURE[t]{\includegraphics[angle=90,width=14cm]{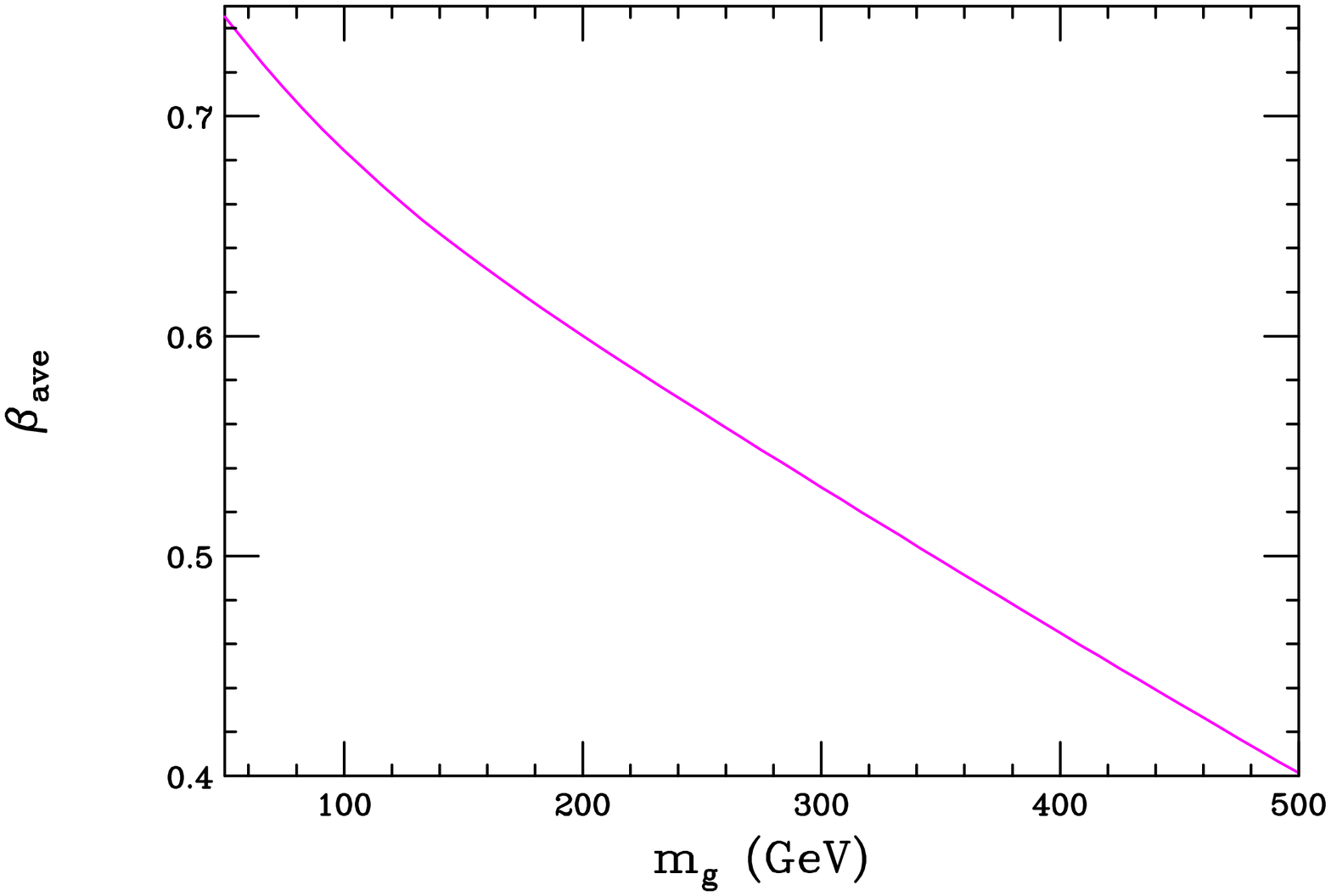}\label{fig:speed}
  \caption{Average speed of a gluino produced at the Tevatron.}}

\subsection{Neutral $R$-hadrons}\label{subsec:neutral}

The most challenging situation in any scenario with a long-lived gluino is the 
case where the gluino fragments to form a neutral stable hadron which remains 
neutral as it traverses the entire detector. As discussed above, a 
neutral gluino hadron will have a mean interaction length of order $\sim 19$ 
cm in the detector but will deposit only a few hundred MeV during each 
interaction. Given this small amount of 
energy deposition it is unlikely that the 
detector will be able to trigger on this signature without being swamped by 
soft QCD backgrounds. In fact, given the typical requirements for defining a 
jet at the Tevatron and LHC detectors, the pair production of gluinos 
fragmenting into neutral stable hadrons will appear as a rather soft process 
and is, hence, {\it invisible}. 
Thus, to trigger on events signaling gluino pair 
production something extra is required. 

Here we will consider the production of a gluino pair in association with an 
additional jet with sufficiently high-$p_T$ so that it captures the
attention of the trigger. Since the $R$-hadron energy 
deposition is rather soft, the gluino-hadron pair plus jet final 
state will be observed in the monojet channel with the neutral gluino 
hadrons appearing as missing energy. There are three subprocesses which 
contribute to this final state at leading order in QCD: $gg, q\bar q 
\rightarrow \tilde g \tilde g+g$ and $gq(\bar q)\rightarrow \tilde g \tilde g
+q(\bar q)$. In ordinary supersymmetry, squarks play an 
important role as intermediate states in these subprocesses. Here, 
as discussed above, 
in the limit where all the squark masses are ultra-heavy, 
the squarks formally decouple. The cross sections corresponding 
to these subprocesses are then found to be essentially the same as that for 
the $t\bar t+$jet final state apart from the appropriate color factors. The 
matrix elements and kinematics required to calculate the relevant cross 
sections can be found in Refs. {\cite {Kunszt:1979iy,Gunion:1986vb}} which we 
adapt for the present 
analysis.

The best current bounds on excess events in the monojet channel from the 
Tevatron are provided from searches for large extra dimensions
by CDF from Run I {\cite {Acosta:2003tz}
(see also {\cite {Abazov:2003gp}}) with 
a data sample of 84 pb$^{-1}$. CDF requires at least one central jet with 
$E_T \geq 80$ GeV as well as missing transverse energy in excess of 80 GeV; if 
an additional jet is also present it must have $E_T \geq 30$ GeV. To be 
called a jet, an energy cluster must have $E_T \geq 15$ GeV in this analysis 
so that the soft energy deposition arising from the neutral $R$-hadrons
will  not be seen as a jet. CDF observes 284 events passing their cuts
while the  SM Monte Carlo predicts $274\pm 16$ from which a bound of 62
can be placed on the number of events arising from New Physics.

\FIGURE[t]{
\includegraphics[angle=90,width=8.2cm]{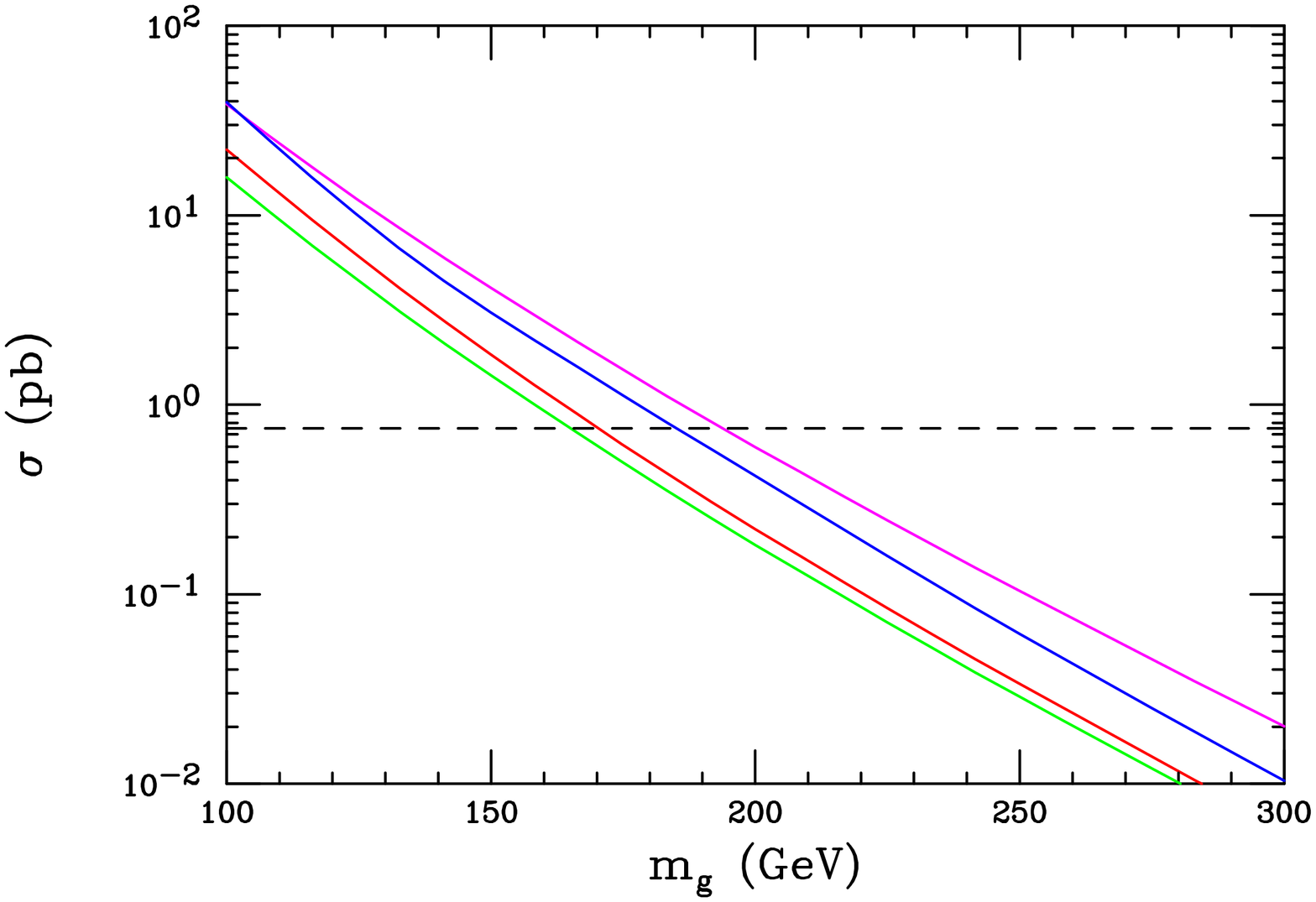}
\includegraphics[angle=90,width=8.2cm]{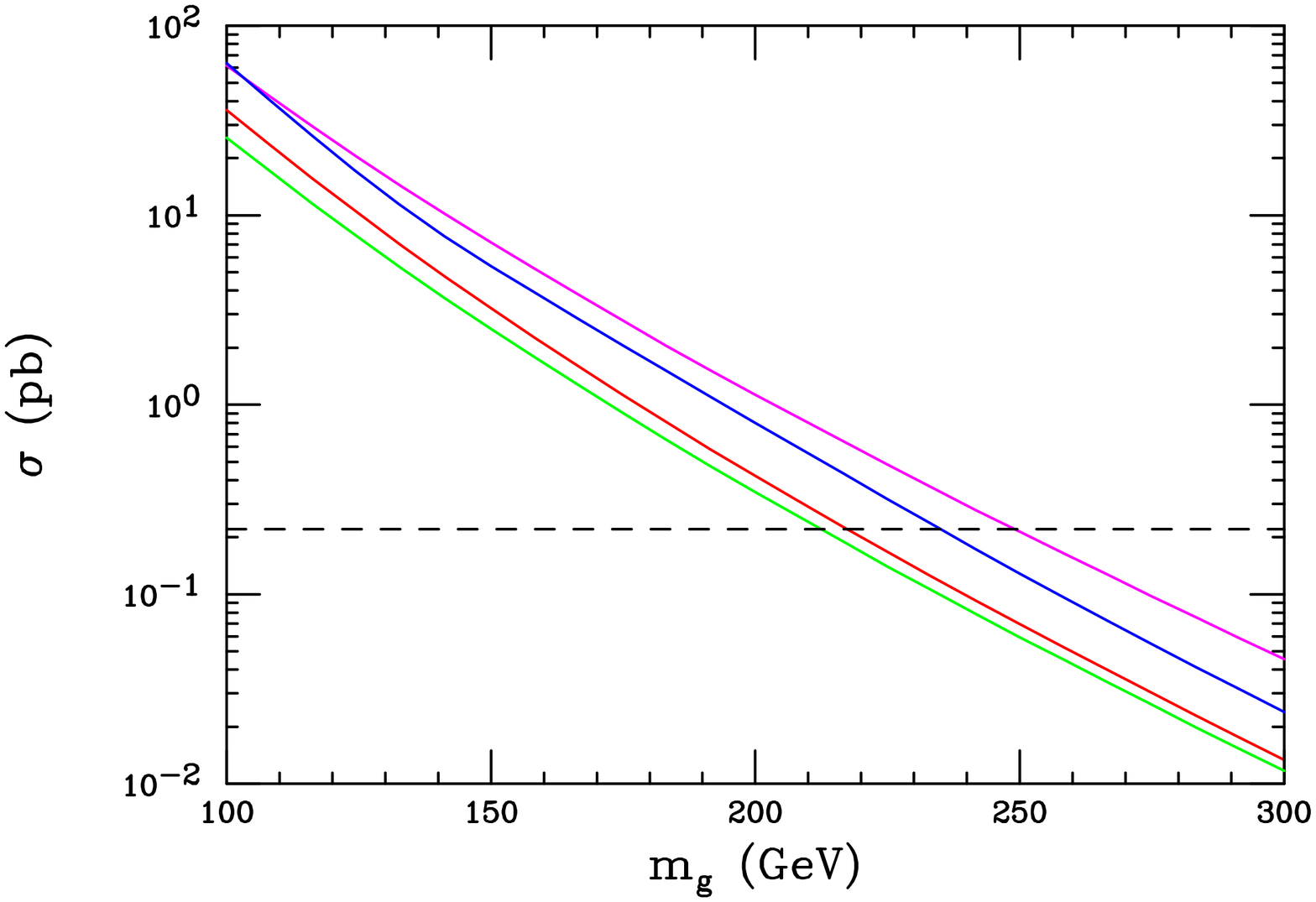}
\includegraphics[angle=90,width=8.2cm]{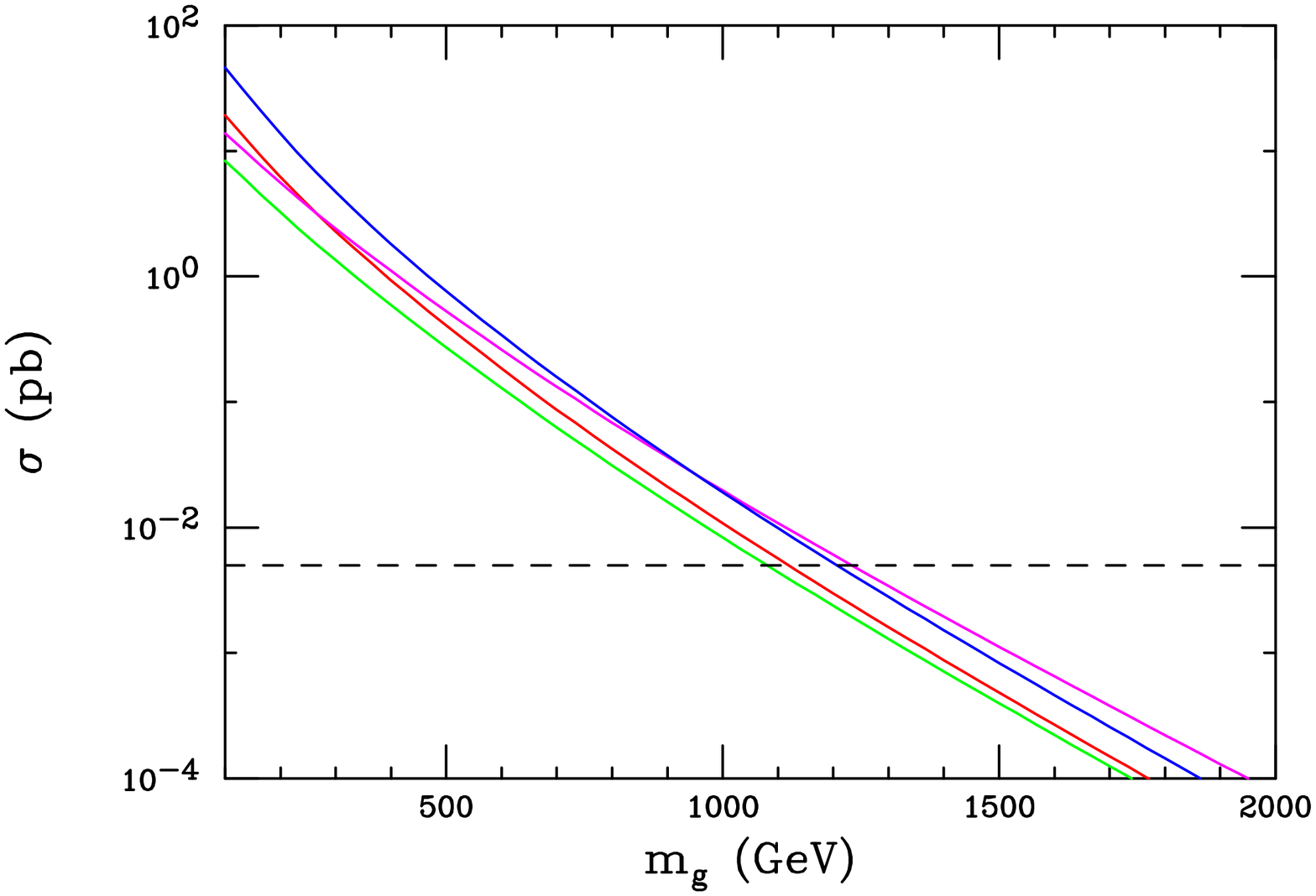}
\label{fig:pairx}\caption{Total cross section for the production of $\tilde g
  \tilde g + {\rm jet}$ at the Tevatron Run I (top), Run II (middle), and LHC
  (bottom). The cross section has been evaluated at four different
  renormalization scales (from top to bottom on the left side): the $p_T$
  of the process (magenta), $m_{\tilde g}$ (blue), the invariant mass of
  the gluino pair (red), the center of mass energy of the parton level
  process (green).}}

In Fig.~\ref{fig:pairx} we show the cross section for 
gluino pairs in association with a jet 
at leading order for the Run I Tevatron ($\sqrt s=1.8$ TeV) as a function of 
the gluino mass employing the CDF cuts. 
Since the calculation is only performed at 
leading order, the renormalization 
scale ambiguities will be rather large; for this reason 
we have evaluated the cross section for several plausible
choices of the renormalization scale which 
leads to a rather broad spread in the cross section predictions. Given the 
CDF bounds, and choosing the scale which leads to the most conservation
cross section, we find that the mass of the gluino, in the case 
where it fragments into a stable neutral hadron, must be in excess of 
$\simeq 170$ GeV from Run I data. 
For Run II we can make an estimate of the 
increased reach in gluino mass if no signal is observed above SM expectations, 
assuming identical detector requirements, by scaling by the square root of the 
integrated luminosity and allowing for the increase of the center of mass 
energy to 1.96 TeV. The results of this analysis are also shown in
Fig.~\ref{fig:pairx}; here we 
see that gluino masses up to $\simeq 215$ GeV or more may be reached with an 
integrated luminosity of 1 fb$^{-1}$. Note that the scale uncertainties remain 
quite large. Looking further ahead, we can repeat this 
analysis for the LHC; here we will require the jet to have $E_T\geq 750$ GeV 
with an identical amount of missing energy to reduce SM backgrounds,
and we will assume an integrated luminosity of $L=100$ fb$^{-1}$. 
The SM backgrounds for this process at the LHC have 
been estimated by Vacavant and Hinchliffe {\cite {Vacavant:2000wz}} 
for the ATLAS detector 
within the context of searches for large extra dimensions. For the present 
analysis, we expect $\simeq 4200$ SM background events to pass our basic cuts. 
Our results are shown in Fig. \ref{fig:pairx} where we see that the 
gluino mass reach is now 
substantially improved to roughly $\simeq 1.1$ TeV, assuming no excess is
observed.

As is well-known, many new physics scenarios can lead to the monojet 
signature. If an excess of events is eventually found at either the Tevatron 
or LHC, is it possible to single out the production of neutral gluino hadrons 
as the source  in any unique way? Here we must recall that while the passage 
of the neutral $R$-hadron through the detector will not be triggered on 
as a jet, the fact that a few hundred MeV of energy will be deposited every 
$\simeq 19$ cm in small `puffs' is a rather unique signature. The idea is
then to 
examine the set of monojet events passing the cuts for additional hadronic 
activity corresponding to two `tracks' of these `puffs' left by the 
hadronic scattering of these neutral gluino hadrons.

Note that this analysis only depends on the fact that the $R$-hadron
is a heavy, strongly interacting particle. Hence, the bounds derived above
apply {\it regardless} of the hadronization pattern. These are thus 
(hadronization) model-independent bounds on the mass of the
gluino. However, if the gluino does hadronize into charged states, the
constraints may be significantly improved.

\subsection{Charged $R$-hadrons}\label{subsec:charged}

If the gluinos fragment predominantly into charged $R$-hadrons
and they do not undergo charge exchange as they interact,
it will be possible to observe them directly
in the detector. In particular,
their time delay (relative to a $\beta=1$ particle)
\cite{Abe:1992vr} or their anomalously high
ionization energy loss \cite{Acosta:2002ju}
as they cross the detector could be observed at the
Tevatron. The bounds in this case will depend
only on the production rate of gluino pairs without any additional
bremsstrahlung.  Recall that the $t$- and $u$-channel diagrams
mediated by squark exchange do not contribute in this model. In
Fig.~\ref{fig:pairproduction} we present
the leading order cross section \cite{Dawson:1983fw}
at the renormalization scale $\mu = 0.2 m_{\tilde g}$,
which is the scale where the LO and NLO
expressions match \cite{Spira:1997cc,Beenakker:1996ch}.

The gluino pairs would be produced with a broad
velocity distribution centered at values of $\beta$ that
range from $\approx 0.7$ for $m=80$ GeV to $\approx 0.4$
for $m=500$ GeV (see Fig. \ref{fig:speed}). In particular,
a significant fraction of them will be produced
with a velocity below $\beta_{max}\approx 0.65$.
For these values of $\beta$, CDF and D0 can distinguish
their time of flight from the tracking chambers to the
muon chambers from that of a $\beta\approx 1$
particle \cite{Abe:1992vr}. In addition, energy loss will
be dominated by ionization, which scales like $1/\beta^2$.
A charged hadron of mass $m\approx  500$ (200) GeV would
be stopped inside the detector if its
initial $\beta$ is below $\beta_{min}\approx 0.25$ (0.35)
(see Section \ref{subsec:prop}).
Therefore, $R$-hadrons with velocity
between $\beta_{max}$ and $\beta_{min}$
can be tracked all the way through the detector,
and their time delay
relative to a $\beta=1$ particle can be measured
and correlated with the energy deposited in the hadronic
calorimeter. For velocities
up to $\beta=0.8$, just the determination of an
anomalously high ionization energy loss (compared to
a $\beta=1$ muon) is enough to establish bounds on the
production rate of these charged $R$-hadrons.

The only published search for stable
charged particles at the Tevatron based on time delay
measures is from CDF, for a small integrated luminosity of
3.54 pb$^{-1}$ \cite{Abe:1992vr}.\footnote{We encourage
CDF and D0 to update their results.}  The results of this
search restrict the total cross section for pair production of
stable particles with unit charge to be smaller than
$\approx 15$ pb for $m = 150$ GeV and
$\approx 5$ pb for $m = 500$ GeV.
In Fig.~\ref{fig:pairproduction} we show that the
gluino pair production cross section intersects the CDF bounds at
$m \approx 200$ GeV.  Scaling by luminosity alone, we would
expect the limit to increase to approximately $m_{\tilde g}
\lsim 270 \gev$ based on time-of-flight measurements with
$100 {\rm pb}^{-1}$ of integrated luminosity collected
at the end of Run I.  For the Tevatron Run II, with a
luminosity of $2$ fb$^{-1}$ this search technique is expected
to cover the region $m_{\tilde g} \lsim 430$ \gev
\cite{Carena:2003yi}.
Note that this is a counting experiment with essentially no physics
background. So, with enough luminosity the reach is simply limited by the
center of mass energy of the collider, as seen from the expected Run II
bound. Note also that the detector signal has a high dependence on the
velocity of the produced stable particle,
and gluinos produced near the kinematic limit will tend to
have the same velocity, regardless of the energy of the collider.
This allows us to estimate the reach at the LHC from time delay searches.
Taking the limit on the number of events to be the same at the LHC as the
Tevatron, we expect
that an integrated luminosity  of 100 fb$^{-1}$
at the LHC could explore the region $m_{\tilde g} \lsim 2.4$ TeV (see
Fig.~\ref{fig:pairproduction}).

CDF has also recently published constraints based on a measure of
anomalous ionization, which they define as events with $dE/dx$
measurements high enough to correspond to $\beta\gamma\le 0.85$.
Looking at Fig.~\ref{fig:speed}, we see that this corresponds
to gluino masses above roughly 100 GeV.
The high $dE/dx$ search technique suffers from backgrounds due to
tracks for which the $dE/dx$ measurement fluctuated high or included
extra ionization from an unreconstructed overlapping particle.
For $\sqrt{s}=1.8$ TeV
and a data sample of 90 pb$^-1$, CDF finds a limit of approximately
0.3 pb for the production cross section of charged
hadrons which do not undergo charge exchange; this bound
is also displayed in Fig.~\ref{fig:pairproduction}.
For the extreme case where the gluino always fragments into a charged
hadron, this implies that its mass should be larger than
$\approx 310$ GeV.  This bound is slightly higher than
our expectations above for results from time-of-flight searches
for heavy stable particles at the end of Run I.  Due to the background
considerations, it is problematic to scale these results for the LHC.

\FIGURE[t]{
\includegraphics[angle=90,width=7cm]{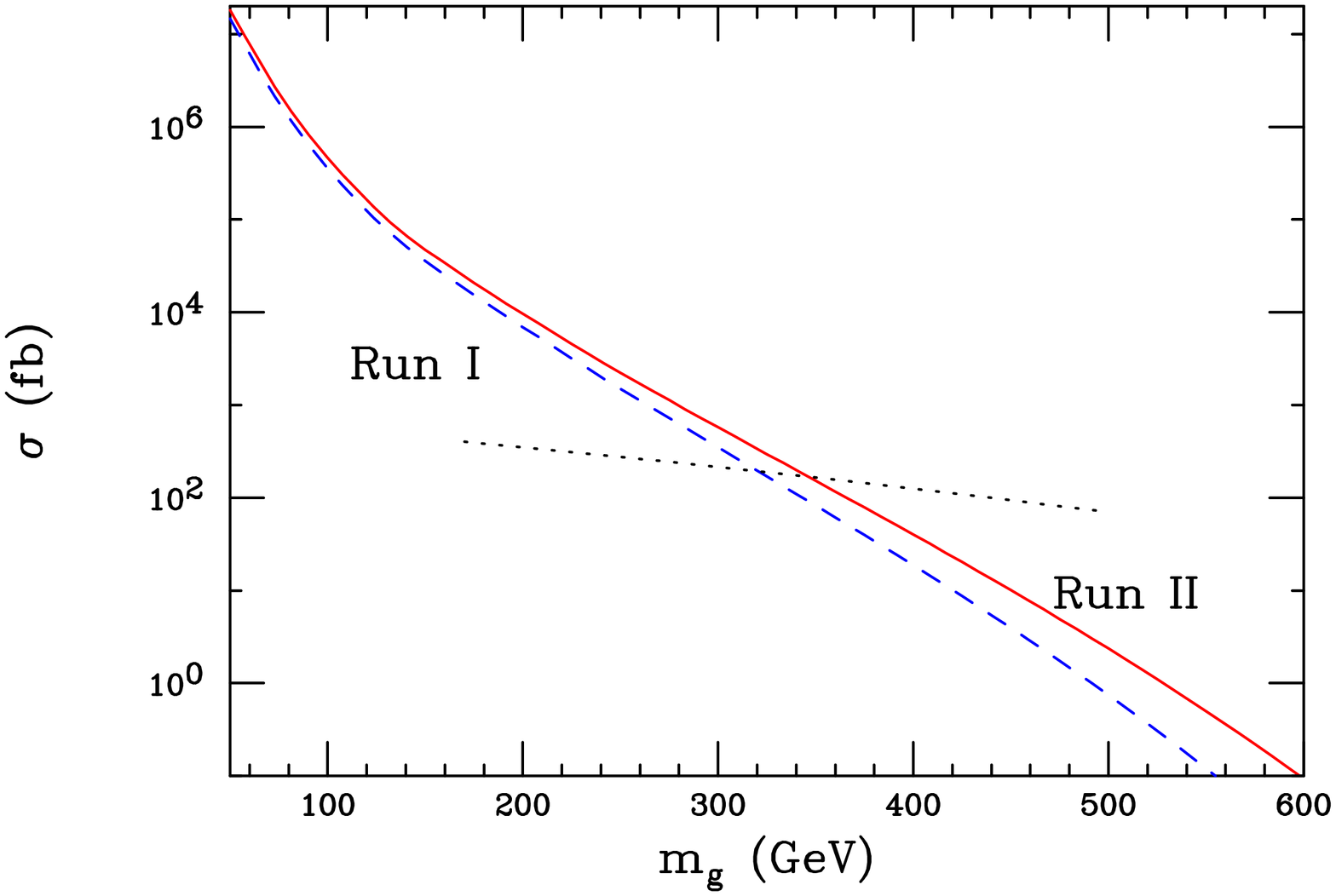}
\includegraphics[angle=90,width=7cm]{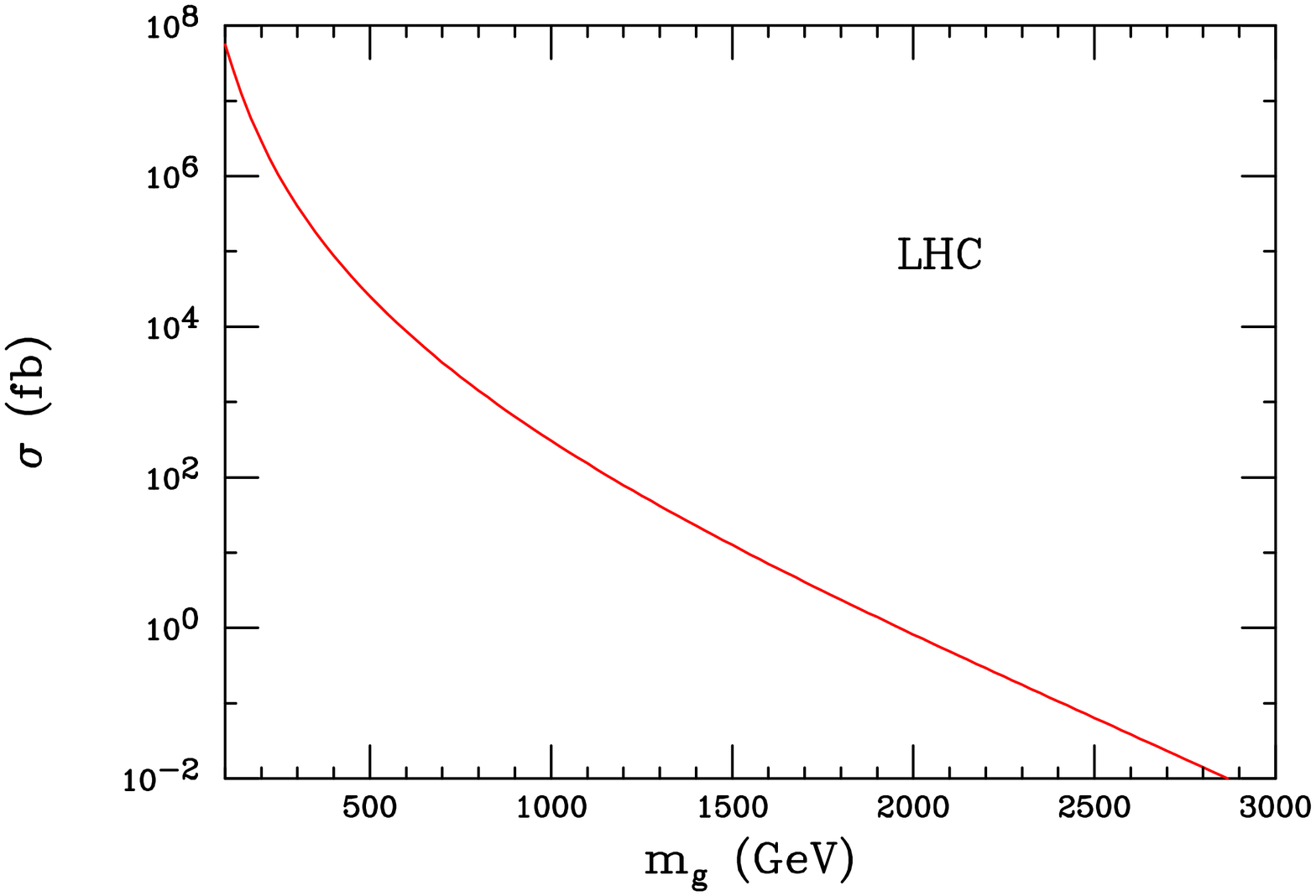}
  \label{fig:pairproduction}\caption{Total cross section for gluino pair
    production at the Tevatron (left) and LHC (right). These are evaluated
    at the renormalization scale $\mu = 0.2 m_{\tilde g}$ suggested by the
    NLO calculation\cite{Spira:1997cc,Beenakker:1996ch}.}}

\subsection{Flippers}\label{subsec:flipper}

An intriguing possibility is that the $R$-hadron can exchange charge
with nuclei as it traverses the detector, so that
the $R$ is ``flipping'' between being neutral and charged.  This
can occur if the $R$-hadron re-fragments after each hadronic interaction
in the detector.

This case can also be directly observed in the detector,
with an efficiency that depends on the frequency of fragmentation
into charged $R$-hadrons versus neutral $R$-hadrons.
The probability rate for this charge exchange
is clearly highly model dependent.  Some models
indicate that this charge exchange will happen if the hadronic
interactions
are dominated by Reggeon exchange, rather than the exchange of Pomerons
or nuclear resonances
\cite{Kraan:2004tz}, and if the mass splitting between the charged
and neutral states is small.
The most difficult signal occurs when
both the produced $R$-hadrons remain neutral most of the time.  In this
case the bound obtained above via the monojet signature will apply.
If the $R$-hadron is charged for a fraction of the time as it crosses the
detector, then the bounds from the previous section will be somewhat
weakened.  However, it is difficult to estimate by how much given the
large uncertainties inherent in modeling the fragmentation.  The best
one can say is that in the case of charge exchange, the constraints on
the gluino mass is bounded by the searches in the two extreme cases
discussed above and lies in the range $170-300 \gev$ from the Tevatron
Run I.

It may be possible to find flipper particles from an offline
analysis of an excess monojet signal. As discussed above, if there is such
an excess, it would be interesting to analyze the data set for
a signal of the underlying physics.  Charge exchange
would result in charged tracks that
stop and start again some distance away.
The presence of these tracks in an excess monojet sample would be a
spectacular signal of a heavy long-lived colored particle.

\section{Cosmic Rays}\label{sec:cosmic}

Let us finally examine the possibility of observing 
the long-lived gluino in non-collider experiments. 
In particular, cosmic rays with energy above 500 TeV 
could produce such particles when they 
strike a nucleon in the upper atmosphere. If that were
the case, the $R$-hadrons could reach the detector 
at IceCube \cite{Spiering:2004dm}
and produce a distinct signature.  Previous studies corresponding
to $R$-hadrons being the source of super-GZK cosmic ray events
can be found in \cite{Chung:1997rz,Berezinsky:2001fy}.

To understand the type of signal that could be expected,
let us consider an initial proton of 
energy $2\times 10^6$ GeV that hits an atmospheric nucleon 
(yielding a center-of-mass energy of $\sqrt{s}=2$ TeV) and creates a pair
of gluinos of 
mass $m_{\tilde g}=200$ GeV. The typical invariant mass of the 
gluinos would be of order 500 GeV, with each gluino 
carrying a total energy  $E_0=2.5\times 10^5$ GeV with 
(equal and opposite) transverse momenta of 100 GeV. The scattering angle 
between the two gluinos would be $\sim 10^{-3}$ rad.

\FIGURE[t]{\includegraphics[angle=0,width=7cm,height=10cm]{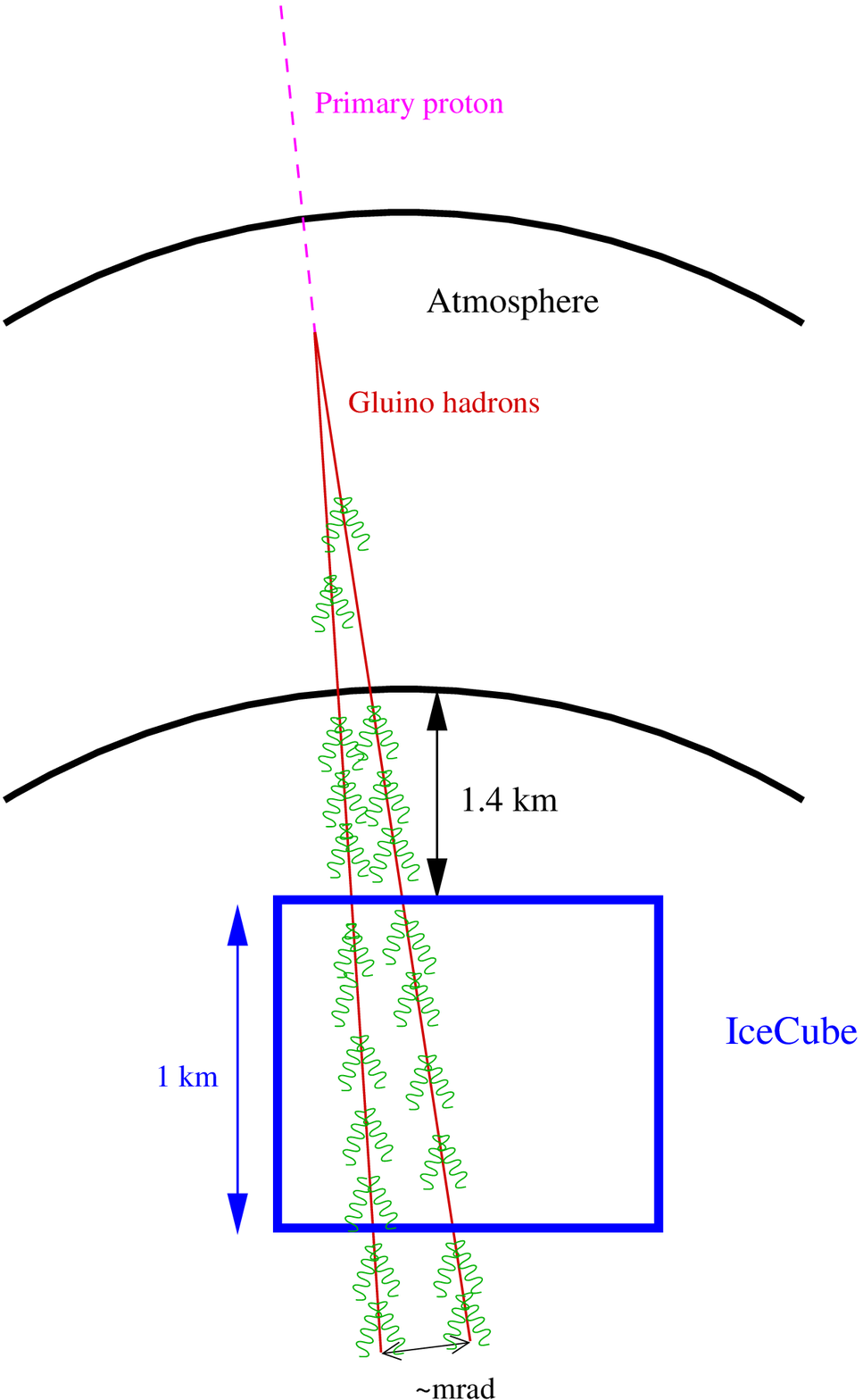}
\label{fig:icefig}\caption{An illustration of the path of $R$-hadron
pairs produced in the atmosphere and striking the IceCube detector.}}

Once hadronized, to reach the top of the IceCube detector the 
$R$-hadrons must cross the atmosphere (an approximate 
depth of $x_a=10^3$ g/cm$^2$) and 1.4 km of ice (a depth of
$x_1=1.4\times 10^5$ g/cm$^2$). In the first
interaction with the atmospheric nucleons, each $R$-hadron
will lose a total energy between 303 GeV 
(case {\it (1)}) and 64 GeV (case
{\rm (2)}).  The two cases, described in section \ref{subsec:prop},
represent two models describing the hadronic interactions of the
heavy $R$-hadrons and represent two extreme scenarios.
At these energies
it is possible to approximate the average energy
loss per interaction by $\Delta E= k \gamma$ with 
$k$ between 0.24 GeV (case {\it (1)}) and 0.055 GeV 
(case {\it (2)}). 
For the interaction length in ice (as well as air) we take 
$\lambda_T(R)=(16/9) \lambda_T(\pi)$, with 
$\lambda_T(\pi)\approx 65$ g/cm$^2$. 
In the linear approximation for the average energy deposited
in each hadronic interaction, $\Delta E$,
the energy of the 
$R$-hadron when it reaches a depth $x$ is 
\begin{gather}
E(x)\approx E_0 e^{-\frac{k x}{ m \lambda_T}}\;.
\end{gather}
Therefore, in the atmosphere each $R$-hadron would deposit 
between 2.7 and 0.6 TeV. This energy would be  
homogeneously distributed, 
and the two jets would reach the ground
separated by approximately $\sim 100$ m. It is unclear 
if such an anomalous profile could be detected in 
air shower experiments (note that this signal would
come {\it inside} a $1.8\times 10^5$ GeV hadronic event).

The $R$-hadrons will then reach the top of the IceCube detector,
which is 1.4 m below the Antarctic surface (see Fig.~\ref{fig:icefig}
for an illustration)
each one with an energy $E(x_1)$ between 57 TeV 
(case {\it (1)}) and 178 TeV (case {\it (2)}). The 
IceCube detector then extends to a depth of 2.4 km.
As the $R$-hadrons pass through the 1000 meters of ice 
($\Delta x = 10^5$ g/cm$^2$), the 
two hadrons could be neutral or could be flipping
charge between each hadronic interaction (approximately one 
per meter). Each $R$ would deposit a total energy 
around 37 TeV (case {\it (1)}) or 38 TeV (case {\it (2)}). 
It is amusing that both cases give a similar amount of 
total energy deposition in the detector: in case {\it (1)} $R$ deposits
a larger fraction of energy per hadronic interaction, 
but it reaches IceCube with less energy than in 
case {\it (2)}. If the hadron is predominantly charged 
there would be $\simeq 200$ GeV of extra energy 
deposited in the form of ionization. 
Although such a pair of $R$ hadrons, separated by 
100 or 150 meters, would be detectable, a
detailed simulation would be needed to see if they 
can be distinguished from a typical muon bundle in 
an air shower core. 

We can now estimate how many of these 
events per year could be expected at IceCube. As a
cosmic proton of energy $E$ enters the atmosphere, 
it will {\it always} interact with a nucleon. If the 
total cross section to produce a pair of gluinos is 
$\sigma^{\tilde g\tilde g}_{pN}$, a fraction 
$\sigma^{\tilde g\tilde g}_{pN}/
\sigma^{Tot}_{pN}$ of these cosmic rays will produce 
the pair of $R$-hadrons. In our estimate we will take the approximation
\cite{Hagiwara:2002fs} of $\sigma(E)^{Tot}_{pN} = 50\;{\rm mb}$.
For $\sigma^{\tilde g\tilde g}_{pN}$ 
we will use the partonic cross sections given in 
\cite{Dawson:1983fw} with the scale $\mu=0.2 m_{\tilde g}$ 
suggested by the next to leading order 
calculation in \cite{Spira:1997cc,Beenakker:1996ch}.

The flux of downgoing protons from zenith angles of
$\cos\theta > 0.5$ can be estimated as \cite{Hagiwara:2002fs} 
${\rm d}F/{\rm d}E=f_1 E^{-2.7}$ 
(year km$^2$ GeV)$^{-1}$ for $E\le 10^7$ GeV and 
${\rm d}F/{\rm d}E= f_2 E^{-3.0}$ 
(year km$^2$ GeV)$^{-1}$ for $E\ge 10^7$ GeV, where 
$f_1\approx 3.0\times 10^{18}$, 
$f_2\approx 3.7\times 10^{20}$, and $E$ is expressed 
in GeV. In addition, we must consider the
possibility that the gluino pair is created in a
second (or later) collision, after the initial proton
has already scattered off a nucleon and retained a fraction of
its initial energy. To be definite, we will consider
that at these energies the proton loses approximately $\sim16\%$ 
of its energy in each hadronic collision (we obtain this 
value from case {\it (1)} in \cite{Baer:1998pg}), and will 
neglect the possibility that the gluinos are produced
by the scattering of secondary particles. This 
effect introduces an additional factor of 
$\sum_{i=0} 0.84^{1.7 i} = 3.9$ in the flux at 
$E < 10^7$ GeV and a similar factor of 3.4 at $E > 10^7$ GeV.

The number of events$/{\rm yr}/{\rm km}^2$ can then be 
estimated as
\begin{gather}
N = \int \frac{{\rm d} F}{{\rm d} E} 
\frac{\sigma^{\tilde g\tilde g}_{pN}}{\sigma^{Tot}_{pN}}
\;{\rm d} E\;.
\end{gather}
In Fig. \ref{fig:icecube} we show the number of events that one may 
expect at the IceCube detector for different values of the gluino
mass. For gluino masses above $170$ GeV we obtain $N<1$, although, of 
course, there would be a non-zero 
possibility to observe an isolated gluino pair event.
Although this search reach is inferior to that from hadron colliders,
it could serve as an independent cross-check of the hadron collider
results.

\FIGURE[t]{\includegraphics[angle=90,width=14cm]{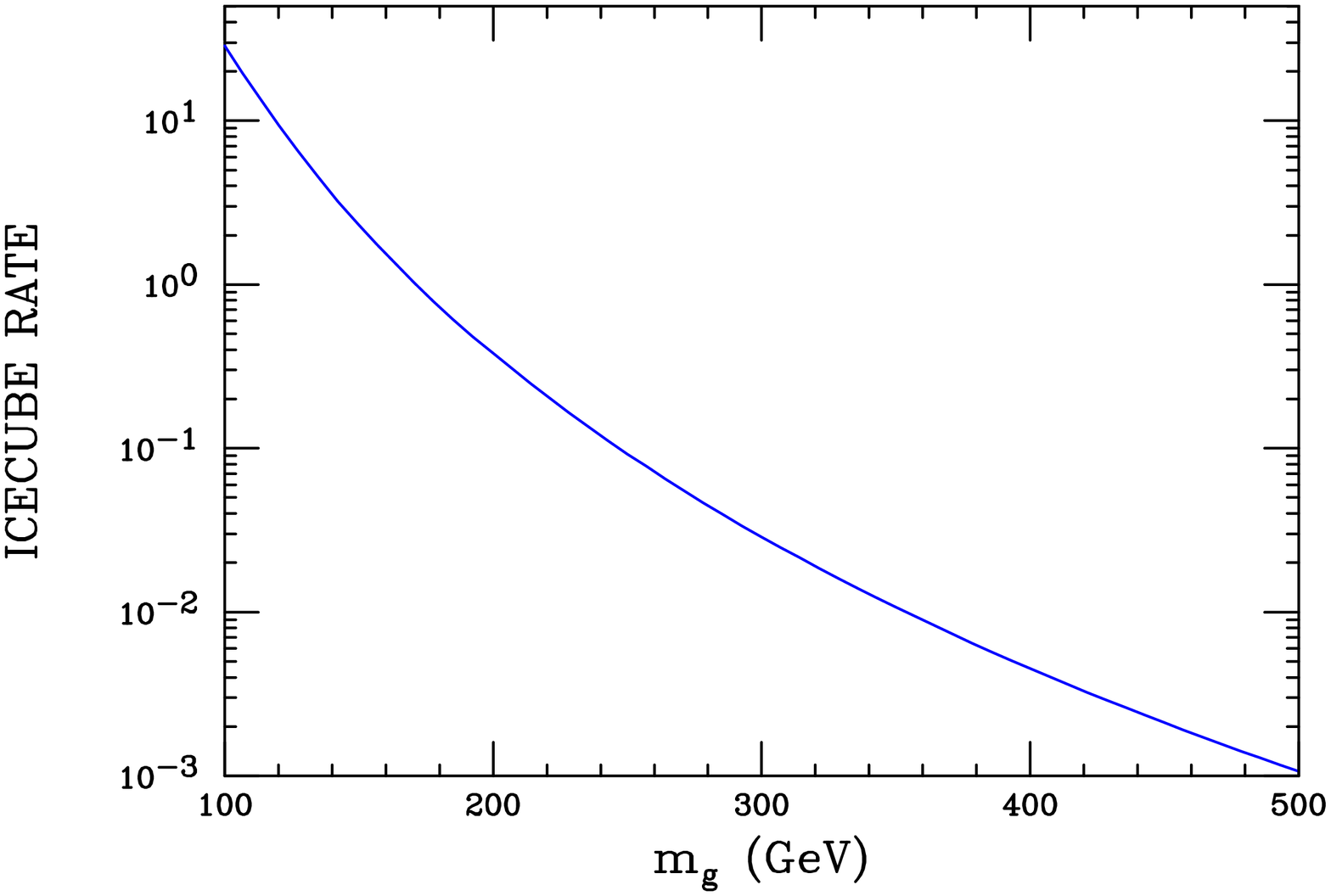}\label{fig:icecube}
  \caption{The number of double gluino events expected per year in IceCube.}}

\section{Summary and Conclusions}\label{sec:conclusion}

In this paper we have examined the experimental signatures for the
production
of gluinos at colliders and in cosmic rays within the Split Supersymmetry
scenario. Unlike in the MSSM, the gluinos in this scenario are relatively
long-lived due to the large value of the squark masses which mediate their
decay. Once this lifetime exceeds a few hundred nanoseconds
the gluino becomes essentially stable as far
as collider and cosmic ray detectors are concerned.

At colliders, gluinos fragment and
form charged or neutral hadrons. It is possible that this fragmentation
may
prefer neutral or charged $R$-hadrons. These hadrons scatter several
times while passing through the detector but only deposit a small
amount of energy during each hadronic interaction.  In the limit
where these hadrons are
charged and remain so as they undergo scattering, they can be discovered
through the conventional stable charged particle search. In this case the
results from the Tevatron Run I place a lower limit on the gluino mass of
$\simeq 310$ GeV which we expect to
increase to $\simeq 430$ GeV for Run II provided no signal
is observed. We can expect that the LHC will extend this into the TeV
range. If the $R$-hadron's charge `flips'
during hadronic interactions in the detector, similar, though reduced,
constraints are found. In the most difficult scenario to observe, the
$R$-hadron is neutral and remains so as it traverses the detector.
The soft energy deposition at each interaction is then seen to be too
small
to be triggered on or pass the cuts defining a jet,
so that ordinary pair production of gluinos will not be
observed. In order to have a reasonable trigger, we considered the
production of a high-$p_T$ jet in association with the gluino pair that
now
appears as missing energy. Such a signature is common to many kinds of new
physics scenarios. From Run I we found that a bound of $\simeq 170$ GeV
can
be placed on the gluino mass in this case, increasing to $\simeq 210$ GeV
at
Run II if no excess events are observed. For the LHC the corresponding
reach is found to be
$\simeq 1.1$ TeV. Note that the $R$-hadrons will produce a
monojet signal regardless of how they hadronize, so the monojet limit is
completely {\it model independent}.

High energy protons in cosmic rays can collide with those in the
atmosphere with sufficient center of mass energy as to produce gluino
pairs
which will have a small opening angle due to the large boost. Due to the
small
energy deposition in each interaction length, these meta-stable gluinos
would
be able to traverse the $\sim 1$ km$^3$ IceCube detector leaving a long
string
of energy depositions which could be observable above backgrounds. We
estimate that the atmospheric gluino pair rate is sufficiently large that
IceCube will be able to probe gluino masses up to 170 GeV. While this is
inside the collider exclusion region, it is an important check in the
case that there are unknown systematics or physics effects that mask the
signatures in colliders.

We have shown that there exist robust constraints on the mass of the
gluino in Split Supersymmetry. Additionally, much of the interesting
region can be probed by the Tevatron Run II, and essentially all of the
interesting region by the LHC.

\bigskip

\noindent
{\bf Note Added}

\bigskip

\noindent 
While this manuscript was in preparation, \cite{Kilian:2004uj} appeared,
where some signatures for long-lived gluino production at the LHC are 
also discussed.

\acknowledgments

The authors would like to thank Nima Arkani-Hamed, John Conway, 
Savas Dimopolous, Tom Gaisser, Francis Halzen, Mike Hildreth, Rich Partridge, 
Greg Landsberg, Enrique Zas,
Chad Davis, and Peter Graham for helpful discussions.

\bibliography{paper}

\end{document}